\documentclass[aip,pof,reprint]{revtex4-2}
\usepackage{graphicx}
\usepackage{epstopdf, epsfig}
\usepackage{color}
\usepackage[dvipsnames]{xcolor}
\usepackage{amssymb}

\newcommand\ufric{u_{\tau}}

\newcommand\Reytau{\mathrm{Re}_{\tau}}

\begin{document}


\title{On the difficulty of determining K\'arm\'an ``constants'' from direct numerical simulations} 



\author{Peter A. Monkewitz}
\email[]{peter.monkewitz@epfl.ch}
\affiliation{\'Ecole Polytechnique F\'ed\'erale de Lausanne (EPFL), CH-1015, Lausanne, Switzerland}


\date{\today}

\begin{abstract}
The difficulty of determining the slope of the famed logarithmic law in the mean velocity profile in wall-bounded turbulent flows, the inverse of the K\'arm\'an ``constant'' $\kappa$, from direct numerical simulations (DNS) is discussed for channel flow. Unusual approaches, as well as the analysis of the standard log-indicator function are considered and analyzed, including the identification of the higher order linear overlap term.
This leads to the conclusion that a definitive determination of the channel flow $\kappa$ from DNS with an uncertainty of, say, 2-3\% will require the residue of the mean stream-wise momentum equation in the simulations to be reduced by at least an order of magnitude.
\end{abstract}

\maketitle 

\section{\label{sec1}Introduction}

Traditionally, the K\'arm\'an parameter $\kappa$ in the logarithmic overlap law of the mean velocity has been determined in simple geometries, such as channel, pipe and flat plate boundary layer flows, from constant portions of the indicator function
\begin{equation}
\label{Xidef}
\Xi_{\mathrm{log}} = y^+\,\mathrm{d}U^+/\mathrm{d}y^+ \equiv Y\,\mathrm{d}U^+/\mathrm{d}Y
\end{equation}
Here, $y^+$ and $U^+$ are the wall normal coordinate and stream-wise mean velocity, inner-scaled with friction velocity $\ufric$ and kinematic viscosity $\nu$, and $Y \equiv y^+/\Reytau$ the outer coordinate, with $\Reytau$ the friction Reynolds number.

Since von K\'arm\'an\cite{vonKarman30,vonKarman31} and Millikan\cite{Millikan}, enormous efforts have been spent to determine the value of $\kappa$ in the mean velocity overlap law
\begin{equation}
\label{UOL}
U^+_{\mathrm{OL}} = (1/\kappa)\,\ln{y^+} + B
\end{equation}
and, nearly 100 years after its introduction, the turbulence community can neither agree on whether $\kappa$ is a universal constant or flow dependent nor on its value. The purpose of the present note is to offer some insight into why these questions have not yet been definitively resolved.

The study focusses on channel flow, for which a multitude of DNS data exist. The selection of data used for this study is summarized in table \ref{TableDNS}, together with the color code used in all the figures.

\begin{table}
\center
\caption{\label{TableDNS} Channel DNS profiles considered in the present study, with the color scheme used in all figures }
\begin{tabular}{l r l l l}
\hline
No. & $\Reytau$ & Reference & color \\
\hline
\#1 & 543 & Lee \& Moser\cite{LM15} & \fcolorbox{white}{green}{$\quad\quad$} \\
\#2 & 640 & Abe et al.\cite{ABE2004} & \fcolorbox{white}{green}{$\quad\quad$} \\
\#3& 944 & Hoyas \& Jim\'enez\cite{HJ06} & \fcolorbox{white}{cyan}{$\quad\quad$} \\
\#4& 1001 & Lee \& Moser\cite{LM15} & \fcolorbox{white}{cyan}{$\quad\quad$} \\
\#5& 1020 & Abe et al.\cite{ABE2004} & \fcolorbox{white}{cyan}{$\quad\quad$} \\
\#6& 1995 & Lee \& Moser\cite{LM15} & \fcolorbox{white}{blue}{$\quad\quad$} \\
\#7& 2003 & Hoyas \& Jim\'enez\cite{HJ06} & \fcolorbox{white}{blue}{$\quad\quad$} \\
\#8& 3996 & Kaneda \& Yamamoto\cite{KY2021} & \fcolorbox{white}{magenta}{$\quad\quad$} \\
\#9& 4179 & Lozano-Dur\'an \& Jimen\'ez\cite{LJ14} & \fcolorbox{white}{magenta}{$\quad\quad$} \\
\#10 & 5186 & Lee \& Moser\cite{LM15} & \fcolorbox{white}{magenta}{$\quad\quad$} \\
\#11 & 7987 & Kaneda \& Yamamoto\cite{KY2021} & \fcolorbox{white}{red}{$\quad\quad$} \\
\#12 & 10049 & Hoyas et al.\cite{HoyasOberlack2022} & \fcolorbox{white}{violet}{$\quad\quad$} \\
\end{tabular}
\end{table}
For channel flow, the stream-wise mean momentum equation is particularly simple
\begin{equation}
\label{MOM}
1 - \mathrm{d}U^+/\mathrm{d}y^+ = Y - \langle uv\rangle^+
\end{equation}
which allows to evaluate $\Xi_{\mathrm{log}}$ from either $U^+$ or $\langle uv\rangle^+$ in sections \ref{sec3} and \ref{sec4}.

Combining equations (\ref{UOL}) and (\ref{MOM}) readily leads to
\begin{equation}
\label{uvOL}
\langle uv\rangle^+_{\mathrm{OL}} = Y - 1 + (\kappa\,y^+)^{-1} \,,
\end{equation}
which is the overlap between the 2-term inner and outer expansions of $\langle uv\rangle^+$ up to order $\mathcal{O}(\Reytau)^{-1}$ (see for instance \citet{AfzalY73} for an early analysis). This is most easily seen by expressing the large $y^+$ expansion of the inner Reynolds stress, up to $\mathcal{O}(\Reytau^{-1})$, $\langle uv\rangle^+_{\mathrm{in}}(y^+\gg 1) = 1 - (\kappa y^+)^{-1} + ... - \Reytau^{-1}[ y^+ + ...]$
and its outer counterpart $\langle uv\rangle^+_{\mathrm{out}} = 1 - Y - \Reytau^{-1} [(\kappa Y)^{-1} + \kappa^{-1}]$ in terms of the intermediate variable\cite{KC81} $\eta = y^+ \Reytau^{-1/2} = Y \Reytau^{+1/2}$.

\begin{figure}
\center
\includegraphics[width=0.7\textwidth]{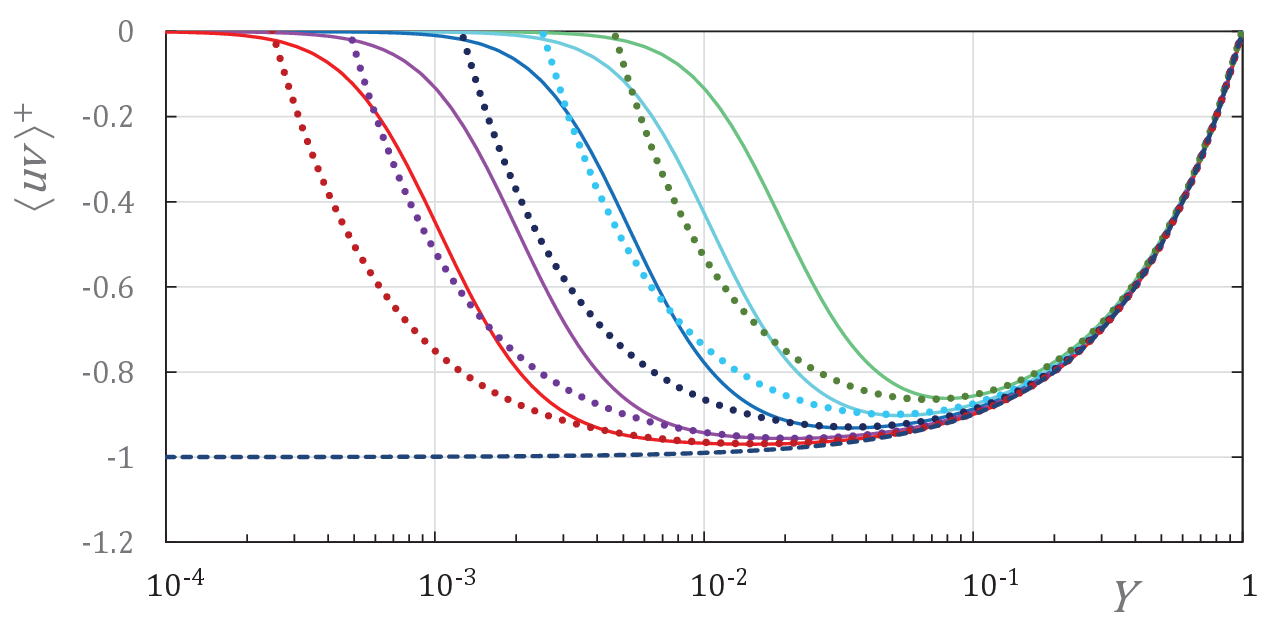}
\caption{\label{Fig1} Reynolds stress from the DNS \#1, 4, 6, 10 and 12 of table \ref{TableDNS} versus outer wall distance $Y$. $\bullet\bullet\bullet$, corresponding fits (\ref{uvOL}) with $\kappa = 0.417$. Black $- - -$: limiting Reynolds stress $(-1 + Y)$ at $\Reytau = \infty$.}
\end{figure}

Realizing that the inner-outer overlap of $U^+$ and $\langle uv\rangle^+$ must coincide and that $\langle uv\rangle^+_{\mathrm{OL}}$ must be located around its minimum, Sreenivasan, known to his friends as Sreeni, deduced the scaling of the minimum Reynolds stress in several landmark publications, \citet{Sreeni87,Sreeni89} and \citet[][fig. 4]{Sreeni97} as
\begin{equation}
\label{minuv}
\langle uv\rangle^+_{\mathrm{min}} = -1 + (3.1 \pm 0.1)\,\Reytau^{-1/2}
\end{equation}

From equation (\ref{uvOL}), one immediately deduces the location and value of the minimum Reynolds stress as
\begin{equation}
\label{minuv2}
y^+_{\mathrm{minRS}} = \left(\frac{\Reytau}{\kappa}\right)^{1/2} \quad\mathrm{and}\quad
\langle uv\rangle^+_{\mathrm{min}} = -1 + \frac{2}{(\kappa\,\Reytau)^{1/2}}
\end{equation}
Comparing equations (\ref{minuv}) and (\ref{minuv2}) shows that Sreeni's estimate of $\kappa = 0.416\,\pm 6.5\%$, obtained from noisy Reynolds stress minima was amazingly close to present day estimates (see for instance \citet{MN2023}, henceforth referred to as ``MN2023'', and references therein).

The above analysis of the Reynolds stress minimum is supported by the channel DNS of figure \ref{Fig1} ``within plotting accuracy'', except for the lowest $\Reytau$ of 543. However, as will be demonstrated in the following, this is not sufficient to ``nail down'' $\kappa$ to better than $\pm 0.5\%$, which is believed necessary to definitively settle the question of its flow dependence.

\section{\label{sec2}The near-wall region}

\begin{figure}
\center
\includegraphics[width=0.8\textwidth]{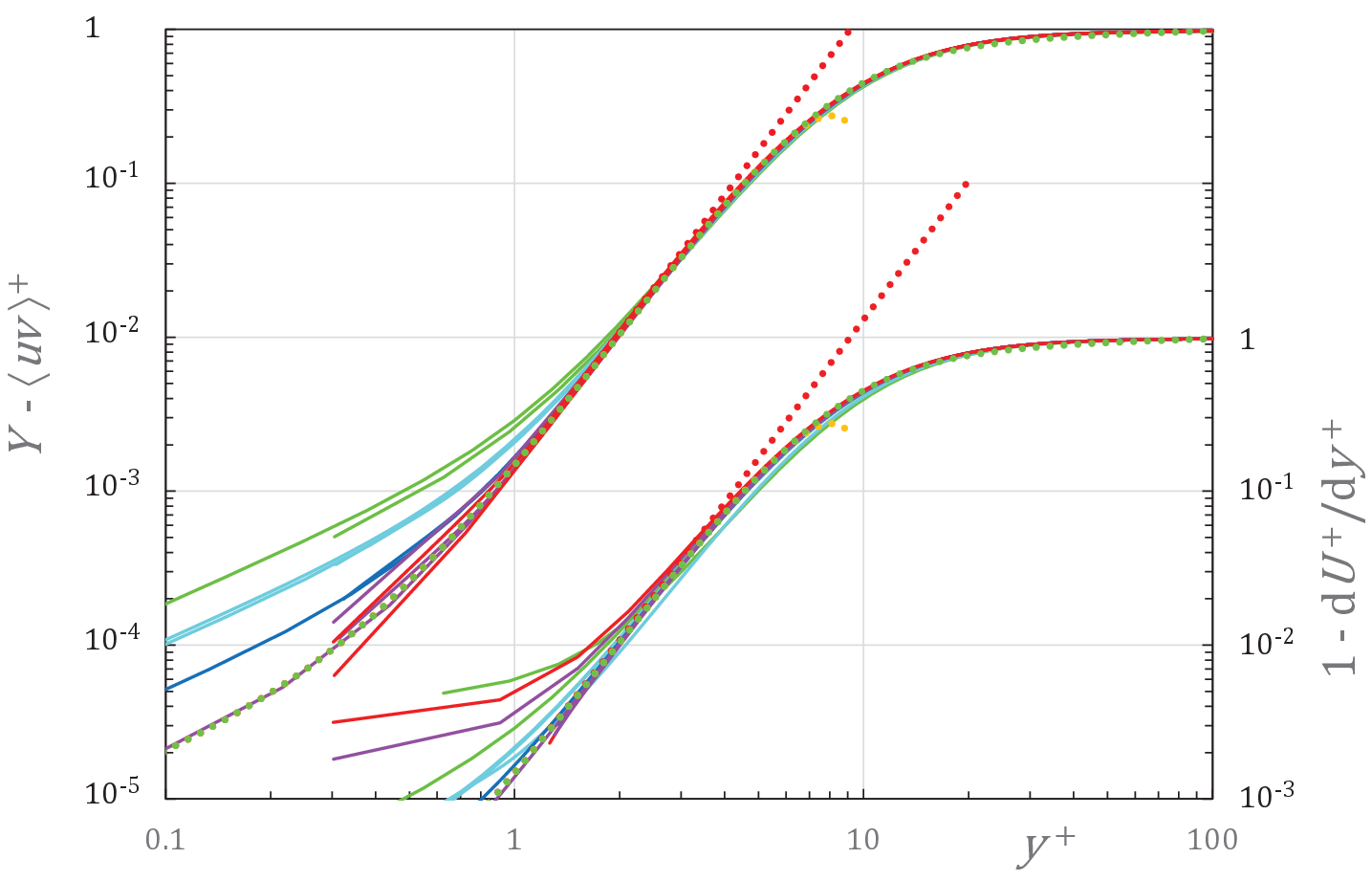}
\caption{\label{Fig2} Comparison of $[Y - \langle uv\rangle^+]$ (left vertical axis) and $[1 - \mathrm{d}U^+/\mathrm{d}y^+]$ (right vertical axis) versus $y^+$, obtained from all the DNS of table \ref{TableDNS}. Red $\bullet\bullet\bullet$, 1st two terms of Taylor expansion (\ref{Taylor}), evaluated for profile \#10; Orange $\bullet\bullet\bullet$, corresponding 3-term expansion; Aqua $\bullet\bullet\bullet$, Pad\'e approximant (\ref{Pade}).}
\end{figure}

In a first step, the consistency of the Taylor expansions of $[Y - \langle uv\rangle^+]$ and of $[1 - \mathrm{d}U^+/\mathrm{d}y^+]$ is examined in figure \ref{Fig2} for all the DNS of table \ref{TableDNS}. The majority of the profiles are seen to be well fitted by the Taylor expansion, which must be anti-symmetric in $y^+$.
\begin{eqnarray}
\label{Taylor}
Y - \langle uv\rangle^+ &&\equiv 1 - \frac{\mathrm{d}U^+}{\mathrm{d}y^+} = \frac{y^+}{\Reytau} + (t_3\,y^+)^3 - (t_5\,y^+)^5 + \mathrm{H.O.T.} \\
&&\mathrm{with}\quad t_3 = 0.109\pm 3\%, \quad t_5 = 0.104\pm10\% \nonumber
\end{eqnarray}
The above coefficient $t_3$ is identical to the one found by \citet{MonkNagib2015} in the zero pressure gradient turbulent boundary layer, but differs from the 0.147 given by \citet{Panton97}. The difference stems from the fact that the Taylor expansion of Panton's Reynolds stress law contains a $(y^+)^4$ term which violates the antisymmetry of $[Y - \langle uv\rangle^+]$ about $y^+ = 0$.

As seen in figure \ref{Fig2}, only the profiles \#2 and \#5 fall significantly below the fit (\ref{Taylor}) for
$[1 - \mathrm{d}U^+/\mathrm{d}y^+]$,
but in view of the following cannot be dismissed outright as outliers. Note also that below $y^+ \approxeq 2$, $[Y - \langle uv\rangle^+ ]$ is dominated by the exact linear term.

Figure \ref{Fig2} also shows, that the two-term Taylor expansion (\ref{Taylor}) provides an excellent fit of the data up to around $y^+ \approx 5$. This range can be extended to beyond 10 by moving, in the spirit of the mean velocity Musker fit\cite{Musker79}, to the Pad\'e approximant
\begin{eqnarray}
\label{Pade}
Y - \langle uv\rangle^+ &&\equiv 1 - \frac{\mathrm{d}U^+}{\mathrm{d}y^+} = \frac{y^+}{\Reytau} + \frac{(t_3\,y^+)^3}{1 + 0.5\,(t_3\,y^+)^2 + (t_3\,y^+)^3} \\
&&\mathrm{with}\quad t_3 = 0.109 \quad \text{as in equ. (\ref{Taylor})}\nonumber
\end{eqnarray}
Equation (\ref{Pade}) may even serve as a rough composite fit, noting however that its asymptotic expansion for $y^+ \gg 1$ is $1 - (2\times 0.109\,y^+)^{-1} + \,\ldots\,\,$, corresponding to a K\'arm\'an parameter of $\kappa = 0.218$ .

\section{\label{sec3}Mean velocity overlap parameters evaluated by fitting asymptotic expansions to $[1 - \mathrm{d}U^+/\mathrm{d}y^+]$ and $[Y - \langle uv\rangle^+]$}

\begin{figure}
\center
\includegraphics[width=0.8\textwidth]{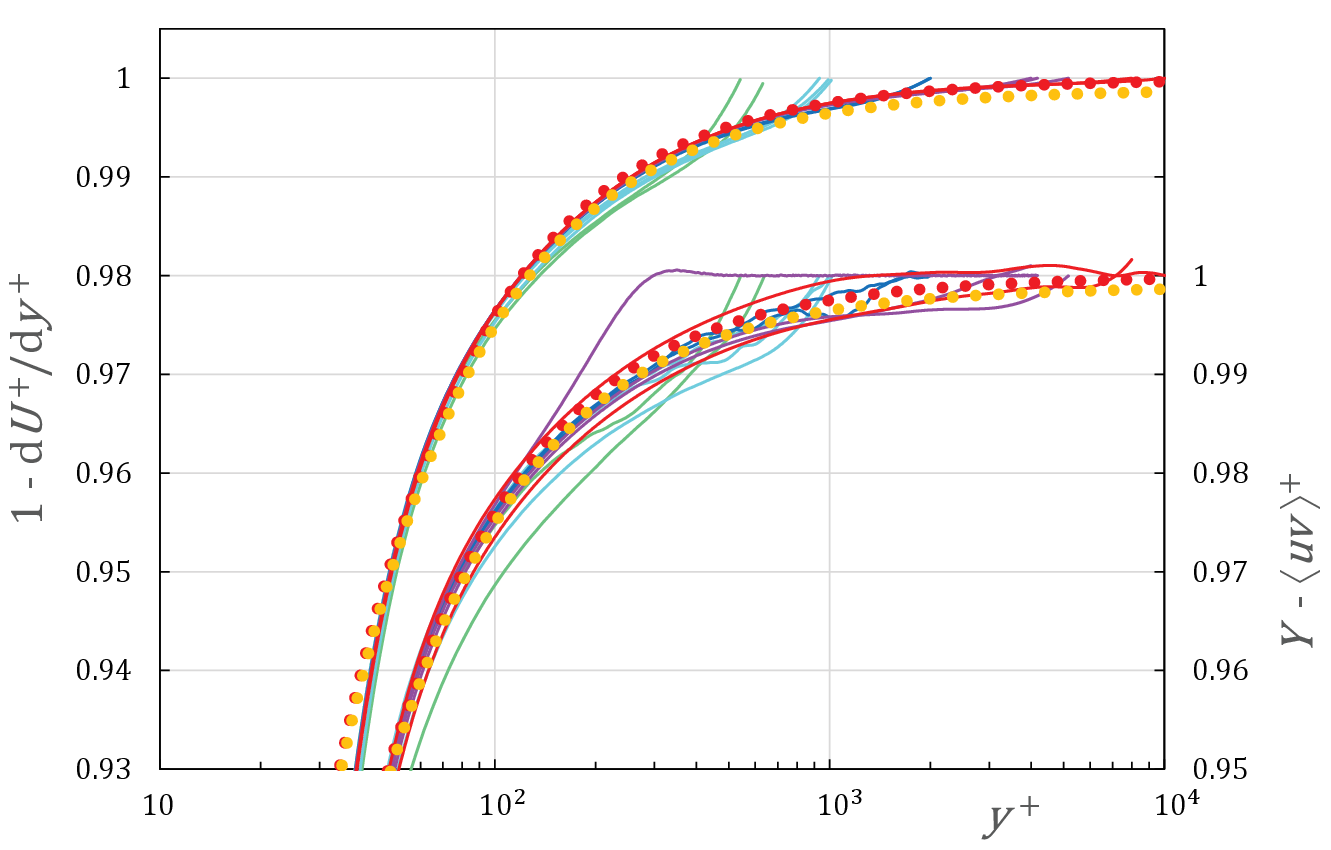}
\caption{\label{Fig3} Enlarged view of the upper part of figure \ref{Fig2} for large $y^+$. Red $\bullet\bullet\bullet$, basic outer asymptotic expansion (1st three terms of equ. \ref{asymp}) with $L=1.15$ and $\kappa = 0.417$, for $\Reytau = 10^4$;
Orange $\bullet\bullet\bullet$, same for $\Reytau = 10^3$.}
\end{figure}

After concluding in section \ref{sec2}, that the Taylor expansions of $[Y - \langle uv\rangle^+]$ and of $[1 - \mathrm{d}U^+/\mathrm{d}y^+]$, extracted from the channel DNS of table \ref{TableDNS} are, apart from minor question marks, reasonably consistent, the more important question of the K\'arm\'an parameter is approached. It can be extracted from either $[1 - \mathrm{d}U^+/\mathrm{d}y^+]$ or $[Y - \langle uv\rangle^+]$, which both have the asymptotic expansion for large $y^+$
\begin{equation}\label{asymp}
1 - \mathrm{d}U^+/\mathrm{d}y^+ \equiv Y - \langle uv\rangle^+ \rightarrow 1 - \frac{L}{\Reytau} - \frac{1}{\kappa\,y^+} + \mathcal{O}(y^+)^{-3} + \frac{1}{\Reytau}\,\frac{\mathrm{d}W}{\mathrm{d}Y}\quad \mathrm{for} \quad y^+ \gg 1
\end{equation}
where $[L/\Reytau]$ is the $y^+$-derivative of the higher order linear term $L\,Y$ in the overlap of $U^+$, most recently discussed by MN2023, and $W(Y)$ is a suitable wake function.

The asymptotic expansion (\ref{asymp}) up to the third term, with $L=1.15$ and $\kappa = 0.417$, as in MN2023, is shown in figure \ref{Fig3} and is seen to fit $[1 - \mathrm{d}U^+/\mathrm{d}y^+]$ rather well for $y^+ \gtrapprox 200$. In particular, the linear term in the overlap of $U^+$, which leads to the $L/\Reytau$ offset in equation (\ref{asymp}), is clearly visible in the data.

A closer look, however, reveals a problem with fitting equation (\ref{asymp}) to the data:
As $y^+$ is reduced from its centerline value of $\Reytau$, the sum of the first three terms of the asymptotic expansion (\ref{asymp}) is seen to first fall below the present DNS data of $[1 - \mathrm{d}U^+/\mathrm{d}y^+]$ before crossing them at $y^+\approxeq 50$. This crossing prevents the asymptotic expansion (\ref{asymp}) from being continued beyond the $(y^+)^{-1}$ term. To maintain the sum of the first three terms of (\ref{asymp}) above and to the left of the DNS data in figure \ref{Fig3} and to ``leave room'' for the next term $\propto (y^+)^{-3}$, the K\'arm\'an parameter $\kappa$ would have to be increased beyond $\kappa\approx 0.5$, which is beyond most estimates. A possible fix of the problem is to introduce a positive offset in the log-law, which is compatible with its asymptotic character. However, as long as the glaring discrepancy between $[1 - \mathrm{d}U^+/\mathrm{d}y^+]$ and $[Y - \langle uv\rangle^+]$ is not clarified, there is no point in perfecting the asymptotic expansion of $[1 - \mathrm{d}U^+/\mathrm{d}y^+]$.
The above discrepancy is further analyzed in the next section \ref{sec4}, where it is magnified by looking at the log-indicator function (\ref{Xidef}).

\section{\label{sec4}Overlap parameters evaluated from the log indicator function}

After revealing the problems of determining overlap parameters by fitting asymptotic expansions to $[1 - \mathrm{d}U^+/\mathrm{d}y^+]$ and $[Y - \langle uv\rangle^+]$ in section \ref{sec3}, the standard method of looking for constant portions of the log indicator function (\ref{Xidef}), i.e. for the K\'arm\'an ``constant'', is investigated next.

\begin{figure}
\center
\includegraphics[width=0.48\textwidth]{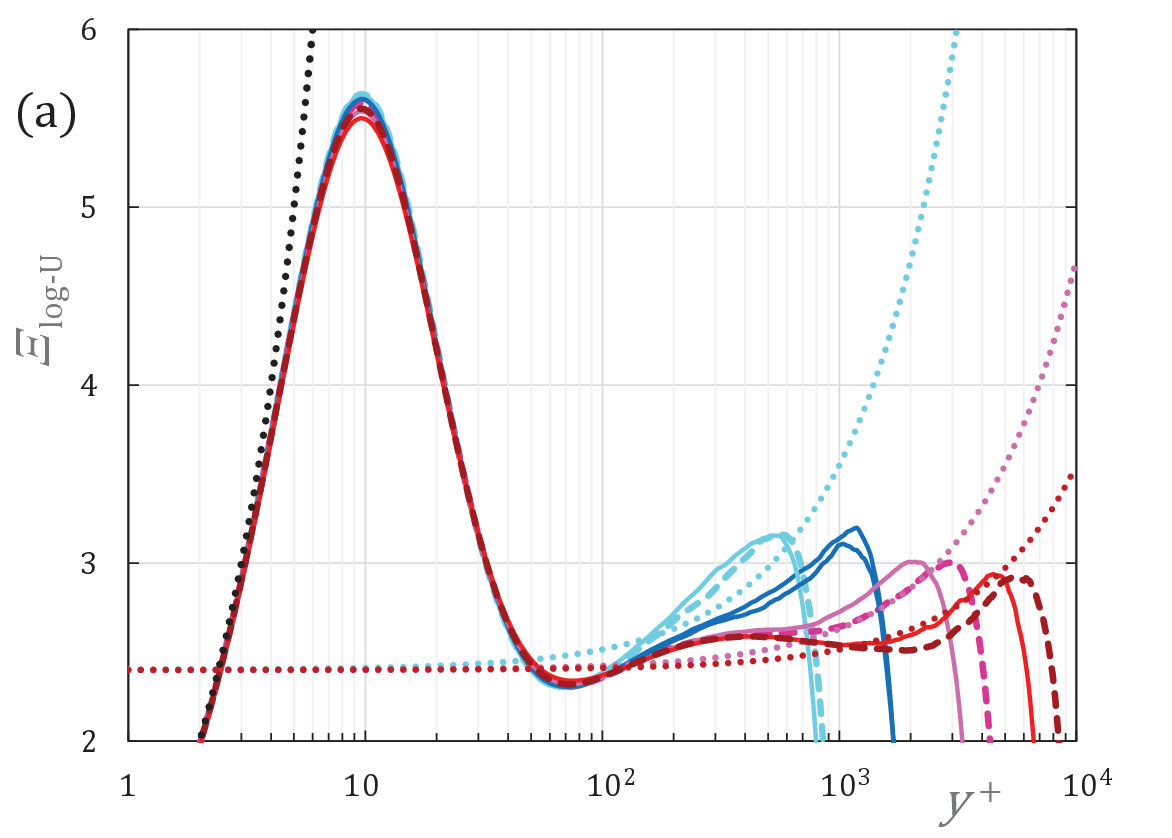}
\includegraphics[width=0.48\textwidth]{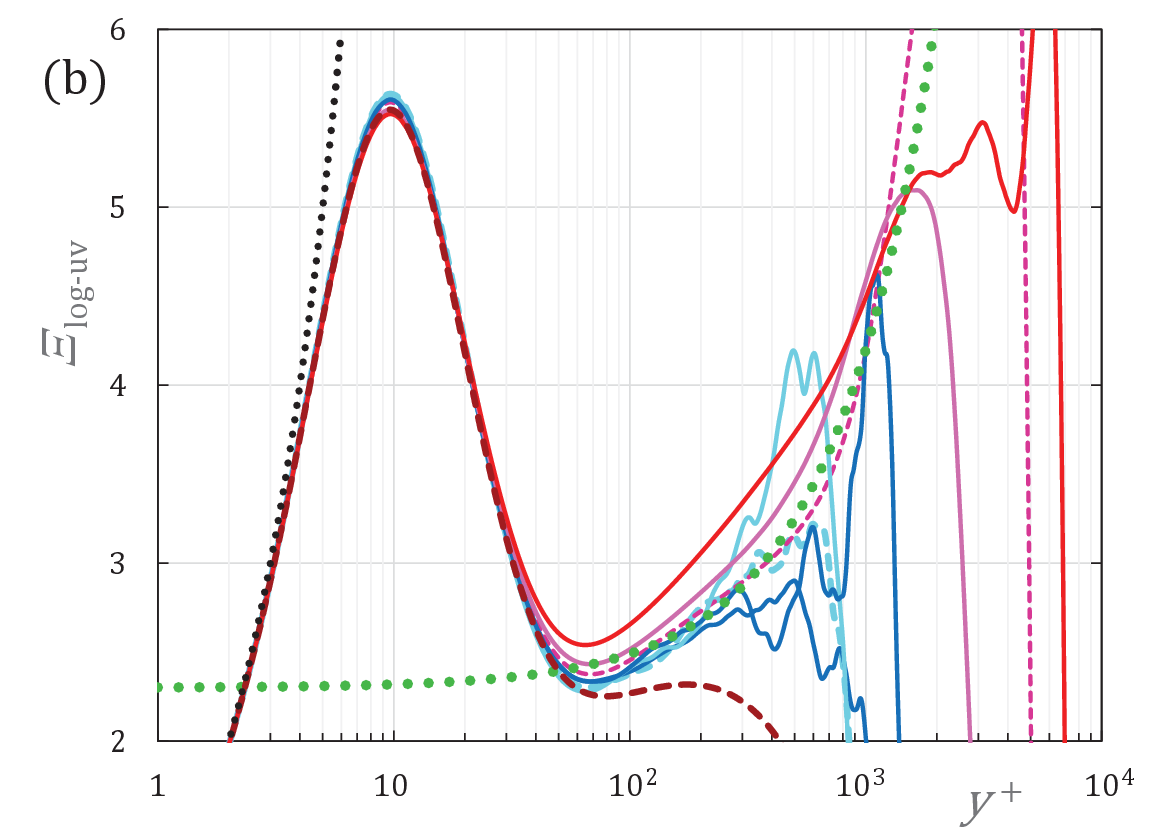}
\caption{\label{Fig5} Log indicator function for DNS \#3, 4, 6, 7, 8, 10, 11, 12 of table \ref{TableDNS} with color code given in the table. Profiles \#4, 10 and 12 are highlighted by broken lines (see text). Black $\cdot\cdot\cdot$, leading Taylor term $y^+$ of $\Xi_{\mathrm{log}}$.
(a) $\Xi_{\mathrm{log-U}} = y^+ \mathrm{d}U^+/\mathrm{d}y^+$;
Aqua, pink and dark red $\cdot\cdot\cdot$, linear overlap term $1.15\,Y$ of MN2023 for $\Reytau = 10^3, 5.10^3$ and $10^4$.
(b) $\Xi_{\mathrm{log-uv}} = y^+[1 - Y + \langle uv\rangle^+]$; Green $\cdot\cdot\cdot$, $\Xi_{\mathrm{log-uv}} = 2.3 + 2.10^{-3}\,y^+$.}
\end{figure}

Both variants of the indicator function, $\Xi_{\mathrm{log-U}} \equiv y^+ \mathrm{d}U^+/\mathrm{d}y^+$ and $\Xi_{\mathrm{log-uv}} \equiv y^+[1 - Y + \langle uv\rangle^+]$, are shown in figures \ref{Fig5}(a) and (b), respectively.
While the indicator functions $\Xi_{\mathrm{log-U}}$ for different $\Reytau$ in figure \ref{Fig5}(a) are clearly more \textit{consistent} than the $\Xi_{\mathrm{log-uv}}$ in figure \ref{Fig5}(b), the \textit{true} channel $\Xi_{\mathrm{log}}$ must be somewhere in between. In other words, extracting a reliable value for $\kappa$ poses problems with both variants.

The indicator $\Xi_{\mathrm{log-U}}(y^+)$ of figure \ref{Fig5}(a) is the one commonly shown in the literature. It consists of an inner-scaled part, independent of $\Reytau$, from which the outer-scaled parts emerge at $Y \approx 0.2$. These outer parts start out approximately linear, before going to zero on the centerline. On the basis of DNS and experimental data, this linear overlap part for channel flow has been fitted in MN2023 by $1.15\,Y$.

Figure \ref{Fig5}(b), on the other hand, shows $\Xi_{\mathrm{log-uv}} = y^+[1 - Y + \langle uv\rangle^+]$ and reveals large differences relative to $\Xi_{\mathrm{log-U}}$, as well as between different DNS. The $\Xi_{\mathrm{log-uv}}$ for different $\Reytau$ and from different authors are seen to feather out already at $y^+ \approxeq 30$.
Furthermore, all but the DNS \#12 approximately follow, beyond $y^+ \approx 80$, the \textit{inner-scaled} linear function $[2.3 + 0.002\,y^+]$, the green dotted line in figure \ref{Fig5}(b). Also, most $\Xi_{\mathrm{log-uv}}$ rise much higher than the corresponding $\Xi_{\mathrm{log-U}}$, before starting to drop to zero beyond $Y \approx 0.4$. The only exception is DNS \#12 where $\Xi_{\mathrm{log-uv}}$ shows no appreciable rise beyond $y^+ \approx 80$, which is most likely due to an unusually small computational box with dimensions $L_x = 2\,\pi h$ and $L_z = \pi h$, which must increase the ``organisation'' of turbulence fluctuations.

Despite the reservations brought up above, it is of interest to further analyze $\Xi_{\mathrm{log-U}}$ of figure \ref{Fig5}(a). In a first step, the fit $L(Y) = 1.15\,Y$ of MN2023 for the outer-scaled part of $\Xi_{\mathrm{log-U}}$ is refined for the present set of DNS profiles to
\begin{equation}
\label{lindelay}
\widehat{L}(Y) = \frac{1.5}{50}\,\ln\{1 + e^{50(Y - 0.09)} \rightarrow 1.5\,Y - 1.35\quad \mathrm{for} \quad e^{50(Y - 0.09)} \gg 1
\end{equation}
which represents a linear function branching off the inner part around $Y = 0.09$. It is noted in passing that, contrary to $L(Y)=1.15\,Y$, $\widehat{L}(Y)$ becomes exponentially small for $Y < 0.09$ and does therefore not have a match in the inner expansion of $\Xi_{\mathrm{log-U}}$.

\begin{figure}
\center
\includegraphics[width=0.8\textwidth]{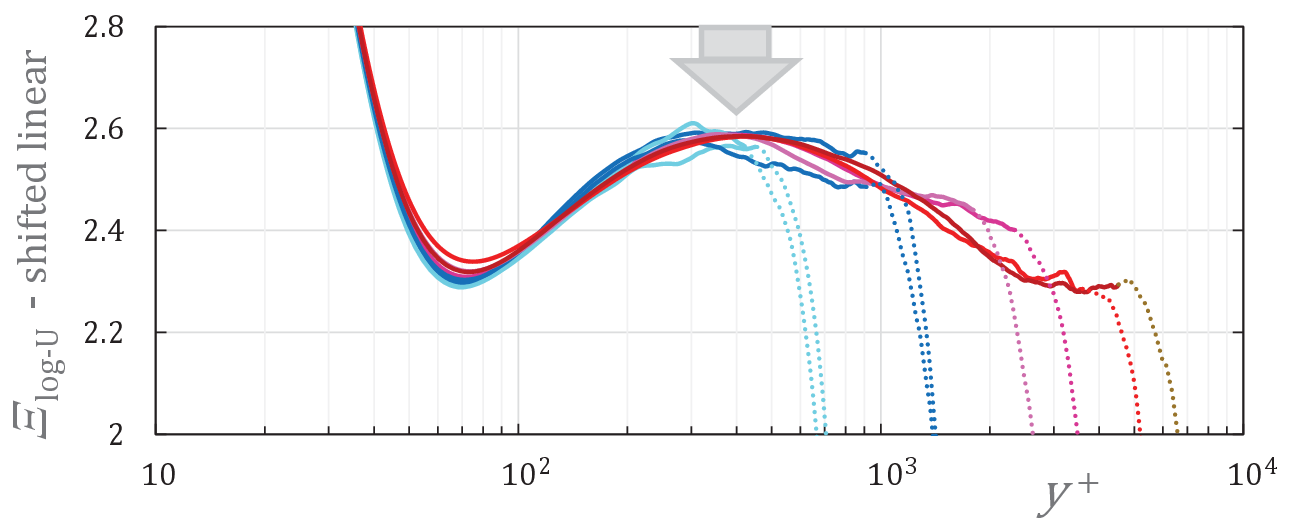}
\caption{\label{Fig6} Indicator function $\Xi_{\mathrm{log-U}}$ as in figure \ref{Fig5}(a), minus the shifted linear function (\ref{lindelay}). The wake part beyond $Y=0.45$ is shown as $\,\,\cdot\cdot\cdot\,\,$.}
\end{figure}

The result of subtracting $\widehat{L}(Y)$ from $\Xi_{\mathrm{log-U}}$ is shown in figure \ref{Fig6} and is seen to produce, for the DNS considered, an excellent collapse of the inner parts of $\Xi_{\mathrm{log-U}}$, up to $Y = 0.45$, where the wake contributions (indicated by dotted lines) become appreciable.
The main result is that, over the range of available $\Reytau$'s, this inner part oscillates between 2.3 ($\kappa = 0.435$) and 2.6 ($\kappa = 0.385$) without an indication on whether it will eventually settle on a constant.

The difference to $\Xi_{\mathrm{log-uv}}$, and the close coincidence of the outer maximum of $[\Xi_{\mathrm{log-U}} - \widehat{L}(Y)]$ at $y^+ \approx 400$ (indicated by the grey arrow in figure \ref{Fig6}) and the emerging outer maximum \cite{monkewitz22} of $\langle uu\rangle^+$, gives rise to the speculation that the spatial oscillations may be due to
a small, $y^+ -\,$dependent bias in the separation between mean and fluctuating stream-wise velocity. Pushing the speculation further, it is noted that, if plotted versus $(y^+)^{1/4}$ (consistent with the $\Reytau^{-1/4}$ scaling of $\langle uu\rangle^+$ \cite{chen_sreeni2021,chen_sreeni2022,chen_sreeni2023,Monkarxiv23}), the period of the spatial oscillation in figure \ref{Fig6} becomes approximately constant and the graph starts to look like the spatial version of a simple under-damped step response in control theory, with the step being the sudden switch-on of the no-slip condition at the wall.

In conclusion, no matter what the correct interpretation of figures \ref{Fig5} and \ref{Fig6} will be, they illustrate why the turbulence community still cannot agree on K\'arm\'an ``constants'', even without considering the unusual indicator functions $\Xi_{\mathrm{log-uv}}$ derived from the Reynolds stress $\langle uv\rangle^+$.

\section{\label{sec5}Conclusions}

The present extensive data analysis, using both standard and non-standard diagnostics, has uncovered several previously unknown or neglected problems with the determination of channel mean velocity overlap parameters, in particular with the slope of the log-law, i.e. the K\'arm\'an parameter. Here, the focus has been on a clear descriptions of the problems and not on proposing still another set of parameters.

In section \ref{sec3} the large differences, beyond $y^+ = \mathcal{O}(10)$, between the computed profiles of $[1 - \mathrm{d}U^+/\mathrm{d}y^+]$ and of $[Y - \langle uv\rangle^+]$ have been documented. Both profiles have allowed to construct a reliable Taylor expansion about the wall. For large $y^+$, on the other hand, fitting an asymptotic expansion to $[1 - \mathrm{d}U^+/\mathrm{d}y^+]$ encounters problems and developing an asymptotic fit of $[Y - \langle uv\rangle^+]$ becomes virtually impossible due to inconsistent data. This is not entirely surprising, as the evaluation of $[1 - \mathrm{d}U^+/\mathrm{d}y^+]$ involves ``only'' the separation of the total stream-wise velocity into mean and fluctuating parts, which appears much less sensitive to the specific implementation of a DNS than determining the Reynolds stress.

Section \ref{sec4}, devoted to the analysis of the log indicator function (equ. \ref{Xidef}), reinforces the conclusions of section \ref{sec3}, that the K\'arm\'an parameter $\kappa$ for channel flow cannot be determined with the desired accuracy from presently available DNS. Worse, all the standard log indicator functions $\Xi_{\mathrm{log-U}} \equiv y^+\,\mathrm{d}U^+/\mathrm{d}y^+$, shown in figure \ref{Fig6}, consistently exhibit, after subtracting their outer-scaled part, an oscillation with a peak-to-peak amplitude of 0.3 , up to the highest available $\Reytau$. This corresponds to $\kappa$ oscillating between 0.385 and 0.435 !

Several interpretations of and speculations on the oscillations of $\Xi_{\mathrm{log-U}}$ may be considered for future study, which will require massive computations. Three possible outcomes are envisioned:
\begin{enumerate}
\item[-] The oscillations are an artefact of the implementation of the DNS and in a ``perfect'' DNS, the inner part of $\Xi_{\mathrm{log}}$ will monotonically approach $\kappa^{-1}$.
\item[-] In a DNS with significantly reduced residue of the mean momentum equation (\ref{MOM}) and $\Xi_{\mathrm{log-U}} \approxeq \Xi_{\mathrm{log-uv}}$, the oscillations in figure \ref{Fig6} become more strongly damped and the different indicators converge to a unique $\kappa^{-1}$, specific to the channel.
\item[-] It is finally conceivable that, even in DNS of extreme quality, which respects $\Xi_{\mathrm{log-U}} \approxeq \Xi_{\mathrm{log-uv}}$ at Reynolds numbers beyond $10^4$, the oscillations persist. Such an outcome would mean that the log law is only an approximate construct, a sacrilege !
\end{enumerate}

To answer these questions and resolve the problems of current DNS highlighted in the present paper, the reduction of the residue of the momentum equation should have absolute priority. In some cases, high Reynolds numbers may also be desirable, but going much beyond a $\Reytau$ of $10^4$ does not appear useful, as the proper identification of higher order terms in the asymptotic expansions of $\langle uv\rangle^+$ and $\mathrm{d}U^+/\mathrm{d}y^+$ requires the computational uncertainty to decrease as $\Reytau^{-1}$.

\begin{acknowledgments}
I first met Sreeni in the eighties, standing around a Helium jet in Yale's Mason lab together with his student Paul Strykowski and discussing absolute instability. Ever since, we have kept in touch and I have always appreciated Sreeni's openness and the priority of physics in all the technical discussions with him.\newline\newline
\end{acknowledgments}

\bibliography{apssamp2}

\providecommand{\noopsort}[1]{}\providecommand{\singleletter}[1]{#1}%
\begin{thebibliography}{23}%
\makeatletter
\providecommand \@ifxundefined [1]{%
 \@ifx{#1\undefined}
}%
\providecommand \@ifnum [1]{%
 \ifnum #1\expandafter \@firstoftwo
 \else \expandafter \@secondoftwo
 \fi
}%
\providecommand \@ifx [1]{%
 \ifx #1\expandafter \@firstoftwo
 \else \expandafter \@secondoftwo
 \fi
}%
\providecommand \natexlab [1]{#1}%
\providecommand \enquote  [1]{``#1''}%
\providecommand \bibnamefont  [1]{#1}%
\providecommand \bibfnamefont [1]{#1}%
\providecommand \citenamefont [1]{#1}%
\providecommand \href@noop [0]{\@secondoftwo}%
\providecommand \href [0]{\begingroup \@sanitize@url \@href}%
\providecommand \@href[1]{\@@startlink{#1}\@@href}%
\providecommand \@@href[1]{\endgroup#1\@@endlink}%
\providecommand \@sanitize@url [0]{\catcode `\\12\catcode `\$12\catcode
  `\&12\catcode `\#12\catcode `\^12\catcode `\_12\catcode `\%12\relax}%
\providecommand \@@startlink[1]{}%
\providecommand \@@endlink[0]{}%
\providecommand \url  [0]{\begingroup\@sanitize@url \@url }%
\providecommand \@url [1]{\endgroup\@href {#1}{\urlprefix }}%
\providecommand \urlprefix  [0]{URL }%
\providecommand \Eprint [0]{\href }%
\providecommand \doibase [0]{https://doi.org/}%
\providecommand \selectlanguage [0]{\@gobble}%
\providecommand \bibinfo  [0]{\@secondoftwo}%
\providecommand \bibfield  [0]{\@secondoftwo}%
\providecommand \translation [1]{[#1]}%
\providecommand \BibitemOpen [0]{}%
\providecommand \bibitemStop [0]{}%
\providecommand \bibitemNoStop [0]{.\EOS\space}%
\providecommand \EOS [0]{\spacefactor3000\relax}%
\providecommand \BibitemShut  [1]{\csname bibitem#1\endcsname}%
\let\auto@bib@innerbib\@empty
\bibitem [{\citenamefont {von K\'arm\'an}(1930{\natexlab{a}})}]{vonKarman30}%
  \BibitemOpen
  \bibfield  {author} {\bibinfo {author} {\bibfnamefont {T.}~\bibnamefont {von
  K\'arm\'an}},\ }\bibfield  {title} {\enquote {\bibinfo {title} {Mechanische
  \"{A}hnlichkeit und {T}urbulenz},}\ }\href@noop {} {\bibfield  {journal}
  {\bibinfo  {journal} {Nachr. Ges. Wiss. G\"ottingen, Math. Phys. Klasse}\
  }\textbf {\bibinfo {volume} {5}},\ \bibinfo {pages} {58--76} (\bibinfo {year}
  {1930}{\natexlab{a}})}\BibitemShut {NoStop}%
\bibitem [{\citenamefont {von K\'arm\'an}(1930{\natexlab{b}})}]{vonKarman31}%
  \BibitemOpen
  \bibfield  {author} {\bibinfo {author} {\bibfnamefont {T.}~\bibnamefont {von
  K\'arm\'an}},\ }\href@noop {} {\enquote {\bibinfo {title} {Mechanical
  similitude and turbulence},}\ }\bibinfo {type} {Tech. Rep.}\ \bibinfo
  {number} {TM 611}\ (\bibinfo  {institution} {NASA},\ \bibinfo {year}
  {1930})\BibitemShut {NoStop}%
\bibitem [{Mil(1938)}]{Millikan}%
  \BibitemOpen
  \href@noop {} {\emph {\bibinfo {title} {Proc. 5th Int. Congr. Appl. Mech.}}}\
  (\bibinfo {year} {1938})\ \bibinfo {note} {wiley, NY.}\BibitemShut {Stop}%
\bibitem [{\citenamefont {Lee}\ and\ \citenamefont {Moser}(2015)}]{LM15}%
  \BibitemOpen
  \bibfield  {author} {\bibinfo {author} {\bibfnamefont {M.}~\bibnamefont
  {Lee}}\ and\ \bibinfo {author} {\bibfnamefont {R.~D.}\ \bibnamefont
  {Moser}},\ }\bibfield  {title} {\enquote {\bibinfo {title} {Direct numerical
  simulation of turbulent channel flow up to ${R}e_\tau = 5200$},}\ }\href@noop
  {} {\bibfield  {journal} {\bibinfo  {journal} {J. Fluid Mech.}\ }\textbf
  {\bibinfo {volume} {774}},\ \bibinfo {pages} {395--415} (\bibinfo {year}
  {2015})}\BibitemShut {NoStop}%
\bibitem [{\citenamefont {Abe}, \citenamefont {Kawamura},\ and\ \citenamefont
  {Matsuo}(2004)}]{ABE2004}%
  \BibitemOpen
  \bibfield  {author} {\bibinfo {author} {\bibfnamefont {H.}~\bibnamefont
  {Abe}}, \bibinfo {author} {\bibfnamefont {H.}~\bibnamefont {Kawamura}},\ and\
  \bibinfo {author} {\bibfnamefont {Y.}~\bibnamefont {Matsuo}},\ }\bibfield
  {title} {\enquote {\bibinfo {title} {Surface heat-flux fluctuations in a
  turbulent channel flow up to $\mathrm{Re}_{\tau}$=1020 with {P}r=0.025 and
  0.71},}\ }\href@noop {} {\bibfield  {journal} {\bibinfo  {journal}
  {International Journal of Heat and Fluid Flow}\ }\textbf {\bibinfo {volume}
  {25}},\ \bibinfo {pages} {404--419} (\bibinfo {year} {2004})}\BibitemShut
  {NoStop}%
\bibitem [{\citenamefont {Hoyas}\ and\ \citenamefont
  {Jim\'{e}nez}(2006)}]{HJ06}%
  \BibitemOpen
  \bibfield  {author} {\bibinfo {author} {\bibfnamefont {S.}~\bibnamefont
  {Hoyas}}\ and\ \bibinfo {author} {\bibfnamefont {J.}~\bibnamefont
  {Jim\'{e}nez}},\ }\bibfield  {title} {\enquote {\bibinfo {title} {Scaling of
  the velocity fluctuations in turbulent channels up to ${R}e_\tau = 2003$},}\
  }\href@noop {} {\bibfield  {journal} {\bibinfo  {journal} {Phys. Fluids}\
  }\textbf {\bibinfo {volume} {18}},\ \bibinfo {pages} {011702} (\bibinfo
  {year} {2006})}\BibitemShut {NoStop}%
\bibitem [{\citenamefont {Kaneda}\ and\ \citenamefont
  {Yamamoto}(2021)}]{KY2021}%
  \BibitemOpen
  \bibfield  {author} {\bibinfo {author} {\bibfnamefont {Y.}~\bibnamefont
  {Kaneda}}\ and\ \bibinfo {author} {\bibfnamefont {Y.}~\bibnamefont
  {Yamamoto}},\ }\bibfield  {title} {\enquote {\bibinfo {title} {Velocity
  gradient statistics in turbulent shear flow: an extension of {K}olmogorov's
  local equilibrium theory},}\ }\href {https://doi.org/10.1017/jfm.2021.815}
  {\bibfield  {journal} {\bibinfo  {journal} {Journal of Fluid Mechanics}\
  }\textbf {\bibinfo {volume} {929}},\ \bibinfo {pages} {A13} (\bibinfo {year}
  {2021})}\BibitemShut {NoStop}%
\bibitem [{\citenamefont {Lozano-Dur\'an}\ and\ \citenamefont
  {Jim\'enez}(2014)}]{LJ14}%
  \BibitemOpen
  \bibfield  {author} {\bibinfo {author} {\bibfnamefont {A.}~\bibnamefont
  {Lozano-Dur\'an}}\ and\ \bibinfo {author} {\bibfnamefont {J.}~\bibnamefont
  {Jim\'enez}},\ }\bibfield  {title} {\enquote {\bibinfo {title} {Effect of the
  computational domain on direct numerical simulations of turbulent channels up
  to ${R}e_\tau = 4200$},}\ }\href@noop {} {\bibfield  {journal} {\bibinfo
  {journal} {Phys. Fluids}\ }\textbf {\bibinfo {volume} {26}},\ \bibinfo
  {pages} {011702} (\bibinfo {year} {2014})}\BibitemShut {NoStop}%
\bibitem [{\citenamefont {Hoyas}\ \emph {et~al.}(2022)\citenamefont {Hoyas},
  \citenamefont {Oberlack}, \citenamefont {Alc\'antara-\'Avila}, \citenamefont
  {Kraheberger},\ and\ \citenamefont {Laux}}]{HoyasOberlack2022}%
  \BibitemOpen
  \bibfield  {author} {\bibinfo {author} {\bibfnamefont {S.}~\bibnamefont
  {Hoyas}}, \bibinfo {author} {\bibfnamefont {M.}~\bibnamefont {Oberlack}},
  \bibinfo {author} {\bibfnamefont {F.}~\bibnamefont {Alc\'antara-\'Avila}},
  \bibinfo {author} {\bibfnamefont {S.~V.}\ \bibnamefont {Kraheberger}},\ and\
  \bibinfo {author} {\bibfnamefont {J.}~\bibnamefont {Laux}},\ }\bibfield
  {title} {\enquote {\bibinfo {title} {Wall turbulence at high friction
  {R}eynolds numbers},}\ }\href@noop {} {\bibfield  {journal} {\bibinfo
  {journal} {Phys. Rev. Fluids}\ }\textbf {\bibinfo {volume} {7}},\ \bibinfo
  {pages} {014602} (\bibinfo {year} {2022})}\BibitemShut {NoStop}%
\bibitem [{\citenamefont {Afzal}\ and\ \citenamefont
  {Yajnik}(1973)}]{AfzalY73}%
  \BibitemOpen
  \bibfield  {author} {\bibinfo {author} {\bibfnamefont {N.}~\bibnamefont
  {Afzal}}\ and\ \bibinfo {author} {\bibfnamefont {K.}~\bibnamefont {Yajnik}},\
  }\bibfield  {title} {\enquote {\bibinfo {title} {Analysis of turbulent pipe
  and channel flows at moderately large reynolds number},}\ }\href
  {https://doi.org/10.1017/S0022112073000546} {\bibfield  {journal} {\bibinfo
  {journal} {Journal of Fluid Mechanics}\ }\textbf {\bibinfo {volume} {61}},\
  \bibinfo {pages} {23–31} (\bibinfo {year} {1973})}\BibitemShut {NoStop}%
\bibitem [{\citenamefont {Kevorkian}\ and\ \citenamefont {Cole}(1981)}]{KC81}%
  \BibitemOpen
  \bibfield  {author} {\bibinfo {author} {\bibfnamefont {J.}~\bibnamefont
  {Kevorkian}}\ and\ \bibinfo {author} {\bibfnamefont {J.~D.}\ \bibnamefont
  {Cole}},\ }\href@noop {} {\emph {\bibinfo {title} {Perturbation methods in
  applied mathematics}}}\ (\bibinfo  {publisher} {Springer},\ \bibinfo {year}
  {1981})\BibitemShut {NoStop}%
\bibitem [{\citenamefont {Sreenivasan}(1987)}]{Sreeni87}%
  \BibitemOpen
  \bibfield  {author} {\bibinfo {author} {\bibfnamefont {K.~R.}\ \bibnamefont
  {Sreenivasan}},\ }\bibfield  {title} {\enquote {\bibinfo {title} {A unified
  view of the origin and morphology of the turbulent boundary layer
  structure},}\ }in\ \href@noop {} {\emph {\bibinfo {booktitle} {Turbulence
  Management and Relaminarization}}},\ \bibinfo {editor} {edited by\ \bibinfo
  {editor} {\bibfnamefont {H.}~\bibnamefont {Liepmann}}\ and\ \bibinfo {editor}
  {\bibfnamefont {R.}~\bibnamefont {Narasimha}}}\ (\bibinfo  {publisher}
  {Springer-Verlag},\ \bibinfo {year} {1987})\BibitemShut {NoStop}%
\bibitem [{\citenamefont {Sreenivasan}(1989)}]{Sreeni89}%
  \BibitemOpen
  \bibfield  {author} {\bibinfo {author} {\bibfnamefont {K.~R.}\ \bibnamefont
  {Sreenivasan}},\ }\bibfield  {title} {\enquote {\bibinfo {title} {The
  turbulent boundary layer},}\ }in\ \href@noop {} {\emph {\bibinfo {booktitle}
  {Frontiers in Experimental Fluid Mechanics}}},\ \bibinfo {editor} {edited by\
  \bibinfo {editor} {\bibfnamefont {M.}~\bibnamefont {{Gad-el-Hak}}}}\
  (\bibinfo  {publisher} {Springer-Verlag},\ \bibinfo {year}
  {1989})\BibitemShut {NoStop}%
\bibitem [{\citenamefont {Sreenivasan}\ and\ \citenamefont
  {Sahay}(1997)}]{Sreeni97}%
  \BibitemOpen
  \bibfield  {author} {\bibinfo {author} {\bibfnamefont {K.~R.}\ \bibnamefont
  {Sreenivasan}}\ and\ \bibinfo {author} {\bibfnamefont {A.}~\bibnamefont
  {Sahay}},\ }\bibfield  {title} {\enquote {\bibinfo {title} {The persistence
  of viscous effects in the overlap region and the mean velocity in turbulent
  pipe and channel flows},}\ }in\ \href@noop {} {\emph {\bibinfo {booktitle}
  {Self-Sustaining Mechanisms of Wall Turbulence}}},\ \bibinfo {editor} {edited
  by\ \bibinfo {editor} {\bibfnamefont {R.~L.}\ \bibnamefont {Panton}}}\
  (\bibinfo  {publisher} {Comp. Mech. Publ., Southampton and Boston},\ \bibinfo
  {year} {1997})\ \bibinfo {note} {also available as
  ar{X}iv:physics.flu-dyn/970801}\BibitemShut {NoStop}%
\bibitem [{\citenamefont {Monkewitz}\ and\ \citenamefont
  {Nagib}(2023)}]{MN2023}%
  \BibitemOpen
  \bibfield  {author} {\bibinfo {author} {\bibfnamefont {P.~A.}\ \bibnamefont
  {Monkewitz}}\ and\ \bibinfo {author} {\bibfnamefont {H.~M.}\ \bibnamefont
  {Nagib}},\ }\bibfield  {title} {\enquote {\bibinfo {title} {The hunt for the
  {K}\'arm\'an ‘constant’ revisited},}\ }\href
  {https://doi.org/10.1017/jfm.2023.448} {\bibfield  {journal} {\bibinfo
  {journal} {Journal of Fluid Mechanics}\ }\textbf {\bibinfo {volume} {967}},\
  \bibinfo {pages} {A15} (\bibinfo {year} {2023})}\BibitemShut {NoStop}%
\bibitem [{\citenamefont {Monkewitz}\ and\ \citenamefont
  {Nagib}(2015)}]{MonkNagib2015}%
  \BibitemOpen
  \bibfield  {author} {\bibinfo {author} {\bibfnamefont {P.~A.}\ \bibnamefont
  {Monkewitz}}\ and\ \bibinfo {author} {\bibfnamefont {H.~M.}\ \bibnamefont
  {Nagib}},\ }\bibfield  {title} {\enquote {\bibinfo {title} {Large {R}eynolds
  number asymptotics of the stream-wise normal stress in {ZPG} turbulent
  boundary layers},}\ }\href@noop {} {\bibfield  {journal} {\bibinfo  {journal}
  {J. Fluid Mech.}\ }\textbf {\bibinfo {volume} {783}},\ \bibinfo {pages}
  {474--503} (\bibinfo {year} {2015})}\BibitemShut {NoStop}%
\bibitem [{\citenamefont {Panton}(1997)}]{Panton97}%
  \BibitemOpen
  \bibfield  {author} {\bibinfo {author} {\bibfnamefont {R.~L.}\ \bibnamefont
  {Panton}},\ }\bibfield  {title} {\enquote {\bibinfo {title} {{A Reynolds
  Stress Function for Wall Layers}},}\ }\href
  {https://doi.org/10.1115/1.2819137} {\bibfield  {journal} {\bibinfo
  {journal} {Journal of Fluids Engineering}\ }\textbf {\bibinfo {volume}
  {119}},\ \bibinfo {pages} {325--330} (\bibinfo {year} {1997})},\ \Eprint
  {https://arxiv.org/abs/https://asmedigitalcollection.asme.org/fluidsengineering/article-pdf/119/2/325/5900842/325\_1.pdf}
  {https://asmedigitalcollection.asme.org/fluidsengineering/article-pdf/119/2/325/5900842/325\_1.pdf}
  \BibitemShut {NoStop}%
\bibitem [{\citenamefont {Musker}(1979)}]{Musker79}%
  \BibitemOpen
  \bibfield  {author} {\bibinfo {author} {\bibfnamefont {A.~J.}\ \bibnamefont
  {Musker}},\ }\bibfield  {title} {\enquote {\bibinfo {title} {Explicit
  expression for the smooth wall velocity distribution in a turbulent boundary
  layer},}\ }\href@noop {} {\bibfield  {journal} {\bibinfo  {journal} {AIAA
  J.}\ }\textbf {\bibinfo {volume} {17}},\ \bibinfo {pages} {655--657}
  (\bibinfo {year} {1979})}\BibitemShut {NoStop}%
\bibitem [{\citenamefont {Monkewitz}(2022)}]{monkewitz22}%
  \BibitemOpen
  \bibfield  {author} {\bibinfo {author} {\bibfnamefont {P.~A.}\ \bibnamefont
  {Monkewitz}},\ }\bibfield  {title} {\enquote {\bibinfo {title} {Asymptotics
  of streamwise {R}eynolds stress in wall turbulence},}\ }\href
  {https://doi.org/10.1017/jfm.2021.924} {\bibfield  {journal} {\bibinfo
  {journal} {Journal of Fluid Mechanics}\ }\textbf {\bibinfo {volume} {931}},\
  \bibinfo {pages} {A18} (\bibinfo {year} {2022})}\BibitemShut {NoStop}%
\bibitem [{\citenamefont {Chen}\ and\ \citenamefont
  {Sreenivasan}(2021)}]{chen_sreeni2021}%
  \BibitemOpen
  \bibfield  {author} {\bibinfo {author} {\bibfnamefont {X.}~\bibnamefont
  {Chen}}\ and\ \bibinfo {author} {\bibfnamefont {K.~R.}\ \bibnamefont
  {Sreenivasan}},\ }\bibfield  {title} {\enquote {\bibinfo {title} {Reynolds
  number scaling of the peak turbulence intensity in wall flows},}\ }\href
  {https://doi.org/10.1017/jfm.2020.991} {\bibfield  {journal} {\bibinfo
  {journal} {Journal of Fluid Mechanics}\ }\textbf {\bibinfo {volume} {908}},\
  \bibinfo {pages} {R3} (\bibinfo {year} {2021})}\BibitemShut {NoStop}%
\bibitem [{\citenamefont {Chen}\ and\ \citenamefont
  {Sreenivasan}(2022)}]{chen_sreeni2022}%
  \BibitemOpen
  \bibfield  {author} {\bibinfo {author} {\bibfnamefont {X.}~\bibnamefont
  {Chen}}\ and\ \bibinfo {author} {\bibfnamefont {K.~R.}\ \bibnamefont
  {Sreenivasan}},\ }\bibfield  {title} {\enquote {\bibinfo {title} {Law of
  bounded dissipation and its consequences in turbulent wall flows},}\ }\href
  {https://doi.org/10.1017/jfm.2021.1052} {\bibfield  {journal} {\bibinfo
  {journal} {Journal of Fluid Mechanics}\ }\textbf {\bibinfo {volume} {933}},\
  \bibinfo {pages} {A20} (\bibinfo {year} {2022})}\BibitemShut {NoStop}%
\bibitem [{\citenamefont {Chen}\ and\ \citenamefont
  {Sreenivasan}(2023)}]{chen_sreeni2023}%
  \BibitemOpen
  \bibfield  {author} {\bibinfo {author} {\bibfnamefont {X.}~\bibnamefont
  {Chen}}\ and\ \bibinfo {author} {\bibfnamefont {K.~R.}\ \bibnamefont
  {Sreenivasan}},\ }\bibfield  {title} {\enquote {\bibinfo {title} {Reynolds
  number asymptotics of wall-turbulence fluctuations},}\ }\href@noop {}
  {\bibfield  {journal} {\bibinfo  {journal} {arXiv}\ }\textbf {\bibinfo
  {volume} {2306.02438}},\ \bibinfo {pages} {to appear in JFM} (\bibinfo {year}
  {2023})}\BibitemShut {NoStop}%
\bibitem [{\citenamefont {Monkewitz}(2023)}]{Monkarxiv23}%
  \BibitemOpen
  \bibfield  {author} {\bibinfo {author} {\bibfnamefont {P.~A.}\ \bibnamefont
  {Monkewitz}},\ }\bibfield  {title} {\enquote {\bibinfo {title} {Reynolds
  number scaling and inner-outer overlap of stream-wise {R}eynolds stress in
  wall turbulence},}\ }\href@noop {} {\bibfield  {journal} {\bibinfo  {journal}
  {arXiv}\ }\textbf {\bibinfo {volume} {2307.00612}} (\bibinfo {year}
  {2023})}\BibitemShut {NoStop}%
\end{thebibliography}%

\end{document}